\pdfoutput=1
\documentclass[prx,twocolumn,showpacs,amsmath,amssymb,floatfix,superscriptaddress]{revtex4-1}

\usepackage{graphicx}  
\usepackage{dcolumn}   
\usepackage{bm}        
\usepackage{amssymb}   
\usepackage{amsmath}
\usepackage{mathrsfs}
\usepackage{epigraph}
\usepackage{braket}
\usepackage{gensymb}
\usepackage{lmodern}
\usepackage{lipsum}
\usepackage{marvosym}
\usepackage{mathdots}
\usepackage{ tipa }
\usepackage{bbold}
\usepackage{dsfont}
\usepackage{esint}
\usepackage{xcolor}
\usepackage{appendix}
\usepackage{natbib}
\usepackage[colorlinks=true, linkcolor=blue, urlcolor=blue,
citecolor=blue]{hyperref}

 \DeclareMathOperator{\Tr}{Tr}

\begin{document}

\title{Robustness of quantized transport through edge states of finite length: Imaging current density in Floquet topological vs. quantum spin and anomalous Hall insulators} 

\author{Utkarsh Bajpai}
\affiliation{Department of Physics and Astronomy, University of Delaware, Newark, DE 19716, USA}
\author{Mark J. H. Ku}
\affiliation{Department of Physics and Astronomy, University of Delaware, Newark, DE 19716, USA}
\affiliation{Dept. of Materials Science and Engineering, University of Delaware, Newark, DE 19716, USA}
\author{Branislav K. Nikoli\'{c}}
\email{bnikolic@udel.edu}
\affiliation{Department of Physics and Astronomy, University of Delaware, Newark, DE 19716, USA}

\begin{abstract}
The theoretical analysis of topological insulators (TIs) has been traditionally focused on infinite homogeneous crystals with band gap in the bulk and nontrivial topology of their wavefunctions, or infinite wires whose boundaries host surface or edge metallic states. Such infinite-length edge states exhibit quantized conductance which is insensitive to edge disorder, as long as it does not break the underlying symmetry or introduces energy scale larger than the bulk gap. However, experimental devices contain finite-size topological region attached to normal metal (NM) leads, which poses a question about how precise is quantization of longitudinal conductance and how electrons transition from topologically trivial NM leads into the edge states. This is particularly pressing issues for recently conjectured two-dimensional (2D) Floquet TI where electrons flow from  time-independent NM leads into time-dependent edge states---the very recent experimental realization [J.~W. McIver {\em et al.}, Nat. Phys. {\bf16}, 38 (2020)] of Floquet TI using graphene irradiated by circularly polarized light did not exhibit either quantized longitudinal or Hall conductance. Here we employ charge conserving solution for Floquet-nonequilibrium Green functions (NEGFs) of irradiated graphene nanoribbon to compute  longitudinal two-terminal conductance, as well as spatial profiles of local current density as electrons propagate from NM leads into the Floquet TI. For comparison, we also compute conductance of graphene-based realization of 2D quantum Hall, quantum anomalous Hall  and quantum spin Hall insulators. Although zero-temperature conductance within the gap of these three conventional time-independent 2D TIs of finite length exhibits small oscillations due to reflections at the NM-lead/2D-TI interface, it remains very close to perfectly quantized plateau at $2e^2/h$ and completely insensitive to edge disorder. This is due to the fact that inside conventional  TIs there is only edge local current density which circumvents any disorder. In contrast, in the case of Floquet TI {\em both} bulk and edge local current densities contribute equally to total current, which leads to longitudinal conductance below the 
expected quantized plateau that is further reduced by edge vacancies.  We propose two experimental schemes to detect {\em coexistence} of bulk and edge current densities within Floquet TI:  ({\em i}) drilling a nanopore in the interior of irradiated region of graphene  will induce backscattering of bulk current density, thereby reducing longitudinal conductance by $\sim 28$\%; ({\em ii}) imaging of magnetic field produced by local current density using diamond NV centers.  
\end{abstract}

\maketitle

\section{Introduction}\label{sec:intro}

The defining property of topological insulators (TIs)~\cite{footnote} is the band gap in the energy spectrum of the bulk material and gapless conducting boundary states. They are edge states in the case of two-dimensional (2D) systems or surfaces states in the case of three-dimensional ones~\cite{Haldane2017}. The paradigmatic cases which gave rise to the main concepts~\cite{Hasan2010,Qi2011} in this field are: ({\em i}) quantum Hall insulator (QHI) in 2D electron gas  which requires an external magnetic field to break the time-reversal invariance and whose edge states are {\em chiral} by allowing spin-unpolarized electron to propagate in only one direction; and ({\em ii})  quantum spin Hall insulator (QSHI)~\cite{Kane2005} which is time-reversal invariant, thereby requiring strong spin-orbit coupling effects instead of external magnetic field, and whose edge states appear in pairs with  different chirality and spin polarization. The last experimentally discovered member of 2D TI family is quantum anomalous Hall insulator (QAHI), which requires both nonzero magnetization to break the time reversal invariance and strong spin-orbit coupling effects, with its edge states allowing only one spin species to flow unidirectionally~\cite{Liu2016}. 

\begin{figure}
	\includegraphics[width=8.5cm]{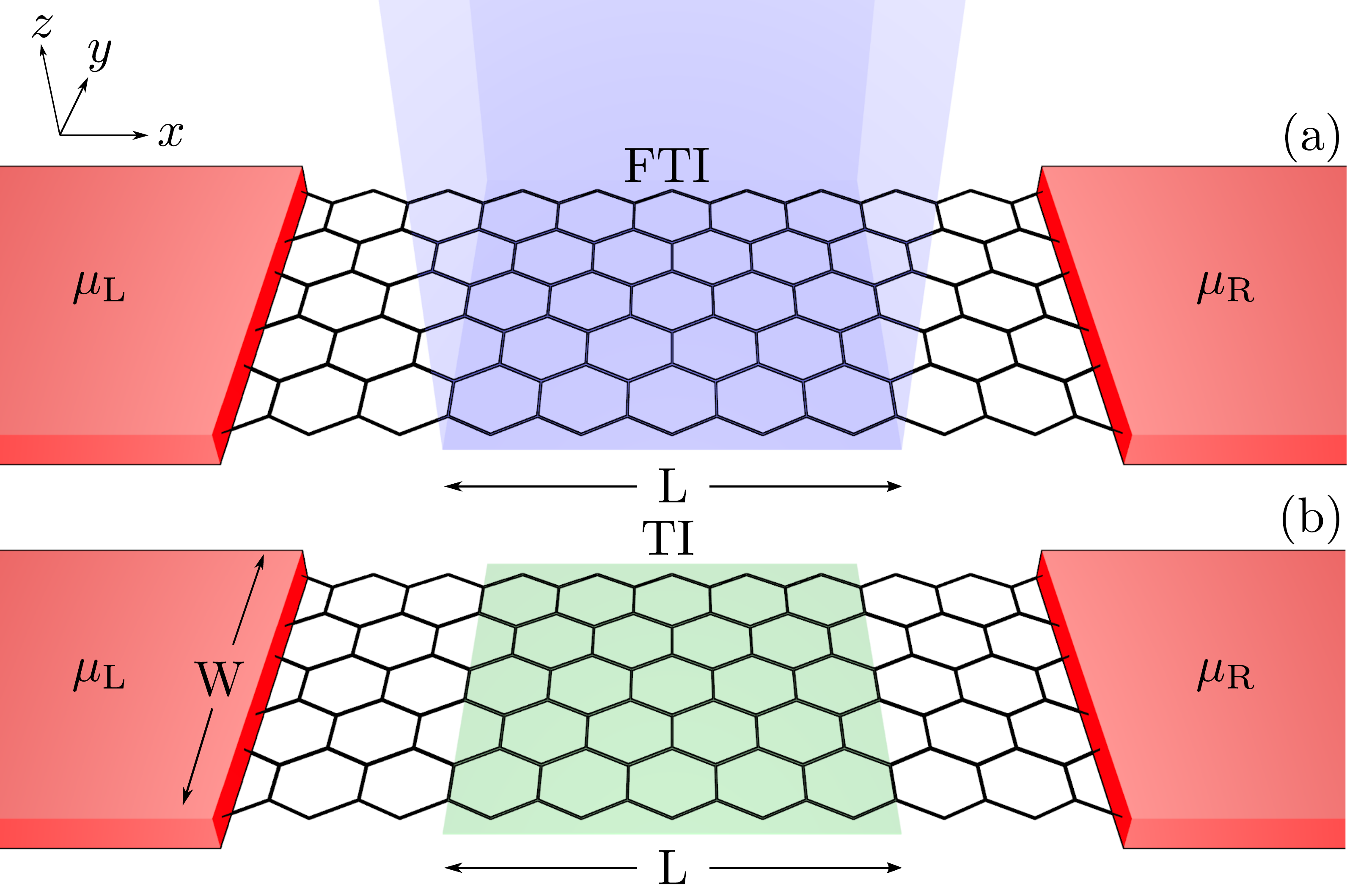}
	\caption{Schematic view of a two-terminal devices where an infinite ZGNR is attached to two macroscopic reservoirs with chemical potentials  $\mu_\mathrm{L}$ and $\mu_\mathrm{R}$ on the left and right, respectively, so that $\mu_\mathrm{L}-\mu_\mathrm{R} = eV_b$ is externally  applied dc bias voltage. In panel (a), the scattering region (blue shaded) in the middle of {\em finite length} $L = 30\sqrt{3}a$ and width $W=29a$ is Floquet TI generated by irradiating~\cite{Oka2009} segment of ZGNR by circularly polarized light of intensity $z$ and frequency $\omega$ [Eq.~\eqref{eq:pphase}]. In panel (b), the scattering region (shaded green)  is quantum Hall, quantum anomalous Hall or quantum spin Hall insulator  with their parameters tuned to produce the same topologically nontrivial band gap $\Delta_g$ [Fig.~\ref{fig:fig2}(a)] in the bulk of all such conventional time-independent TIs.}
	\label{fig:fig1}
\end{figure}

In theoretical analysis, edge states are found as eigenfunctions $\Psi_{k_x}(x,y)=e^{i k_x x}\psi(y)$ of the Hamiltonian of an infinite wire (periodic along the $x$-axis, so that eigenfunctions are labeled by the wavevector $k_x$) made of 2D TIs. The corresponding eigenenergies $\varepsilon(k_x)$ form subbands crossing the band gap~\cite{Halperin1982,MacDonald1984}. The width of the edge states is defined by the spatial region where the probability density is nonzero,  \mbox{$|\psi(y)|^2 \neq 0$}, while decaying  exponentially fast towards the bulk of the wire. Interestingly, their width~\cite{Prada2013,Chang2014} can also depend on the arrangement of atoms along the edge, such as in the case of graphene wires where edge states of QHI and QAHI or QSHI are narrower in the case of zigzag arrangement of carbon atoms along the edge than in the case of their armchair arrangement~\cite{Prada2013,Chang2014,Sheng2017}. In paradigmatic three-dimensional TIs like Bi$_2$Se$_3$, surface states actually have spatial extent of about $\sim 2$ nm~\cite{Chang2015}. 

The zigzag edge, which is employed in devices in Fig.~\ref{fig:fig1}, can also introduce a kink in the subband of edge state~\cite{Sheng2017}, so that subband   intersects with the Fermi energy $E_F$ at $N_\mathrm{R}$ points with positive velocity and $N_\mathrm{L}$ points with negative velocity. However,  only the difference $N_\mathrm{R} - N_\mathrm{L}=|\mathcal{C}|$ is topologically protected according to the {\em bulk-boundary correspondence}~\cite{Hasan2010,Qi2011}. Here  $\mathcal{C}$ is an integer topological invariant (like the Chern number in the case of QHI and QAHI) associated with band structure in the bulk. This makes   electronic transport through edge states of infinite length {\em perfectly quantized in a robust way}~\cite{Buttiker1988}---the zero-temperature two-terminal conductance is $G(E_F)=G_\mathrm{Q} |\mathcal{C}|$  for $E_F$ swept through the bulk band gap and insensitive to both magnetic and nonmagnetic disorder in the case of QHI and QAHI~\cite{Sheng2017}, or only nonmagnetic disorder in the case of QSHI. Although infinite ballistic wires, including those with topologically trivial edge states~\cite{Allen2015,Zhu2017,Marmolejo-Tejada2018}, also exhibit integer $G(E_F)/ G_\mathrm{Q}$, this is easily disrupted by disorder introduced around the edges or even within the bulk~\cite{Marmolejo-Tejada2018}. Here $G_\mathrm{Q}=2e^2/h$ or $G_Q=e^2/h$ is the conductance quantum for spin-degenerate or spin-polarized edges states, respectively.

\begin{figure*}
	\includegraphics[scale=1.1]{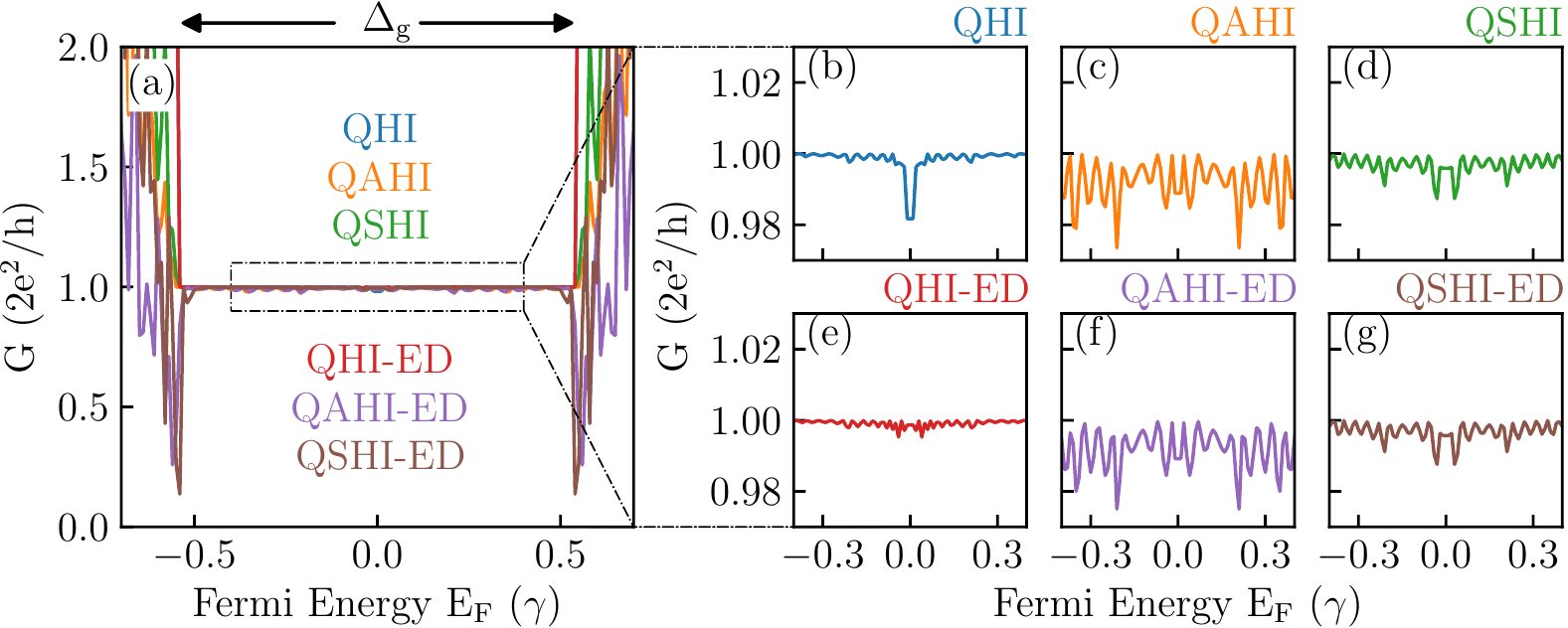}
	\caption{(a) The zero-temperature two-terminal conductance vs. the Fermi energy $E_F$ of device in Fig.~\ref{fig:fig1}(b) where central scattering region hosts {\em finite-length} conventional 2D  time-independent TIs, such as QHI, QAHI and QSHI. The TIs are defined on pristine or edge disordered (denoted by ED) ZGNR due to vacancies illustrated in Figs.~\ref{fig:fig4}(d), ~\ref{fig:fig4}(f) and ~\ref{fig:fig4}(h).  The zoom in of conductance values within the rectangle in panel (a) is shown in: (b)--(d) for pristine ZGNR; and (e)--(g) for edge-disordered ZGNR.  The two NM leads, from which electrons are injected into the topologically protected edge states of finite length with the corresponding local current density profiles shown in Fig.~\ref{fig:fig4}(c)--(h), are also made of ZGNR of the same width as the scattering region [Fig.~\ref{fig:fig1}(b)]. The gap in the bulk of all three 2D TIs is tuned to $\Delta_g = 0.54 \gamma$ and marked in panel (a).}
	\label{fig:fig2}
\end{figure*}

Thus, it has been considered that the key experimental signature of topology in 2D condensed matter is conductance quantization in transport through edge states, which persists even in the presence of disorder as long as it does not break underlying symmetries of the topological phase or generates energy scales larger than the bulk band gap. However, for QHI, QAHI and QSHI of finite length, the zero-temperature longitudinal  conductance $G=I/V_b$, also denoted as `two-terminal' since current $I$ and small bias voltage $V_b$ are measured between the same normal metal (NM) leads,   oscillates in Fig.~\ref{fig:fig2} just below the quantized plateau at $2e^2/h$ while remaining very close to it. We use zigzag graphene nanoribbon (ZGNR) within which 2D TI of finite length [Fig.~\ref{fig:fig1}(b)] is established  using sufficiently large external magnetic field~\cite{Cresti2016}, or additional terms of the Haldane~\cite{Haldane1988} or the Kane-Mele~\cite{Kane2005} models, to generate QHI, QAHI and QSHI, respectively. Their parameters are tuned so that all three examples of conventional time-independent 2D TIs in Fig.~\ref{fig:fig2}(a) have identical topologically nontrivial bulk band gap $\Delta_g$. Even though $G(E_F)$, for $E_F$ swept through bulk band gap $\Delta_g$, is not perfectly quantized in Fig.~\ref{fig:fig2}(a), its oscillations zoomed in Figs.~\ref{fig:fig2}(b)--(g) are insensitive to nonmagnetic edge disorder (ED) introduced in the form of edge vacancies [illustrated in Fig.~\ref{fig:fig4}]. 

It is worth mentioning that imperfectly quantized two-terminal $G(E_F)$ was observed in early experiments on QSHI~\cite{Roth2009}, provoking a lively search for exotic many-body inelastic effects~\cite{Maciejko2009,Tanaka2011,Budich2012,Vayrynen2013,Kainaris2014,Vayrynen2018,Novelli2019} which can circumvent band-topology constraints and introduce  backscattering of electrons as they propagate through edge states. On the other hand, Fig.~\ref{fig:fig2} demonstrates that imperfectly quantized $G(E_F)$ can be due to a much simpler mechanism---backscattering at the NM-lead/TI-region interface.

Lacking perfectly quantized two-terminal conductance $G(E_F)$ as the experimental signature of 2D TI phase of finite length, one can resort to direct imaging of spatial profiles of local current density that should confirm electronic flux confined to a narrow region defined by the edge states. Continuous experimental advances have made this possible, such as by using superconducting interferometry in Josephson junction setup~\cite{Allen2015,Zhu2017} or scanning superconducting quantum interference device (SQUID)~\cite{Nowack2013}. In the latter case, one images  magnetic field produced by the current from  which one can reconstruct the local current density with  $\sim \mu$m spatial resolution~\cite{Nowack2013}. Even higher resolution, with reconstructed images having spatial resolution of $\sim 10$ nm, has been achieved by using scanning tip based on electronic spin of a diamond nitrogen-vacancy (NV)  centers~\cite{Tetienne2017,Chang2017,Ku2020}. Particularly {\em intriguing questions} that such images can answer is how electron flux transitions from topologically trivial NM lead present in every experimental device into the region of 2D TI of finite length where the flux is confined within narrow edge currents, as well as how processes at the NM-lead/2D-TI interface affect the total current and the corresponding conductance. 

\begin{figure}
	\includegraphics[width=\linewidth]{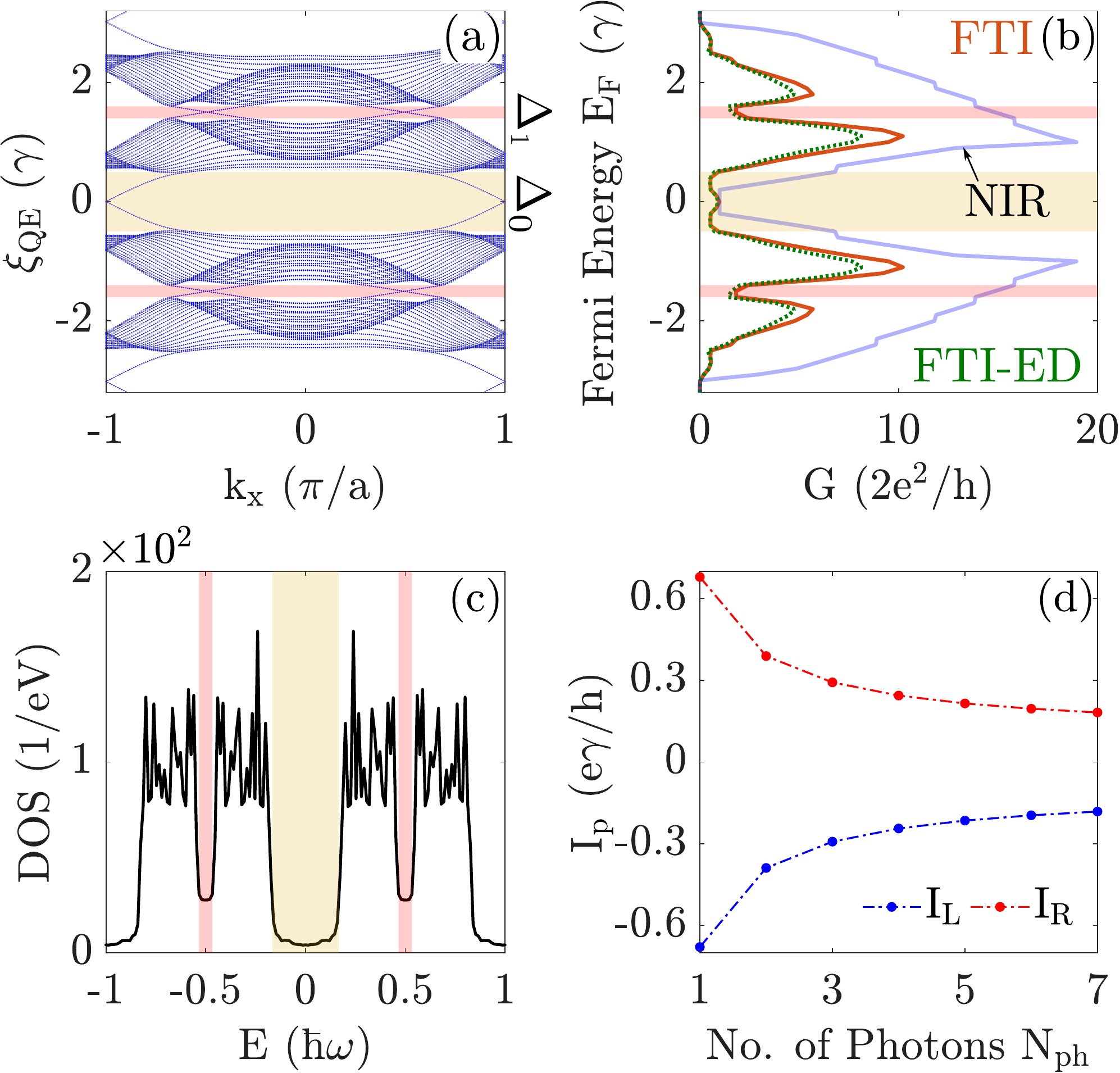}
	\caption{(a) Quasi-energy spectrum $\xi_\mathrm{QE} (k_x)$ for an infinite ZGNR that is irradiated by circularly polarized monochromatic laser light of frequency \mbox{$\hbar \omega = 3 \gamma$} and intensity \mbox{$z=0.5$} over its whole length. The spectrum is obtained by diagonalizing the corresponding Floquet Hamiltonian [Eq.~\eqref{eq:floquet_ham}] truncated to $-N_\mathrm{ph}<n<N_\mathrm{ph}$ Floquet replicas where $N_\mathrm{ph}=7$ is chosen. The yellow shaded region marks the topological gap $\Delta_0$ around $\xi_\mathrm{QE} = 0$ corresponding to CNP, while the red shaded region marks the dynamical  topological gap $\Delta_1$  around  $\xi_\mathrm{QE} = \pm\hbar\omega/2$. (b) The zero-temperature two-terminal conductance vs. $E_F$ (computed using $N_\mathrm{ph} = 7$) of two-terminal device in Fig.~\ref{fig:fig1}(a) whose scattering region is Floquet TI of finite length due to irradiation by circularly polarized light. The pristine irradiated ZGNR is marked by FTI and irradiated edge-disordered ZGNR is marked by FTI-ED. The conductance of an infinite nonirradiated (NIR) pristine ZGNR is also show as a reference. (c) Total DOS for the same device marked by FTI in panel (b). (d) Convergence of lead currents $I_\mathrm{L}$ and $I_\mathrm{R}$ vs. $N_\mathrm{ph}$ at \mbox{$E_F = \hbar\omega/2$ }. }
	\label{fig:fig3}
\end{figure}

Imaging of local current density could also offer new avenue for resolving a {\em crucial issue} for recently conjectured new class of 2D TIs---the so-called Floquet TI~\cite{Oka2009,Lindner2011,Oka2019,Rudner2020}---which is the connection between the Floquet quasi-energy
spectrum and experimentally measurable dc transport properties. The Floquet TI phase arises in 2D electron systems driven out of equilibrium by strong light-matter interaction.  For example,  graphene~\cite{Oka2009,Lindner2011,Oka2019,Rudner2020}, as well as other 2D materials with honeycomb lattice structure like transition-metal dichalcogenides~\cite{Huaman2019}, subject to a spatially uniform and circularly polarized  light are predicted to transmute into Floquet TI with quasi-energy spectrum~\cite{Shirley1965,Sambe1973}. Its multiple gaps share~\cite{Lindner2011} the same topological properties as the band  gap of QAHI described by the Haldane model~\cite{Haldane1988}. This means that the laser induced band gaps, such as $\Delta_0$ in Fig.~\ref{fig:fig3}(a) emerging at the charge neutral point (CNP) of graphene and $\Delta_1$ away from CNP, are crossed by subbands of chiral edge states~\cite{Perez-Piskunow2014,Perez-Piskunow2014a}. The eigenfunctions of these subbands  decay exponentially towards the bulk with a decay length that depends only on the ratio of the laser frequency and its intensity. 

The $\Delta_1$ gaps are called dynamical gaps~\cite{Syzranov2008} and they occur at energy $\hbar \omega/2$ above/below the CNP. They can be reached using experimentally accessible parameters. For example, the very recent experiment~\cite{McIver2020} has been interpreted in terms of creation of a transient Floquet TI by driving graphene flake by \mbox{$500$~fs} laser pulse at a frequency of \mbox{$\omega=46$~THz}, so that the photon energy is \mbox{$\hbar \omega \approx 191$ meV} and its wavelength is \mbox{$\lambda \approx 6.5$ $\mu$m}. However, the experiment of Ref.~\cite{McIver2020} did not observe either quantized longitudinal or transverse (Hall) conductance. Instead, they found that at the peak laser pulse fluence the transverse conductance within $\Delta_0$ gap saturated at plateau around \mbox{$G_{xy}=(1.8 \pm 0.4)e^2/h$}, while no such plateau of $G_{xy}$ was observed within $\Delta_1$ gap.

The calculations of two-terminal [as in Figs.~\ref{fig:fig2} and  ~\ref{fig:fig3}(b)] or multi-terminal conductance typically employ  the Landauer-B\"{u}ttiker setup~\cite{Buttiker1988,Imry1999} depicted in Fig.~\ref{fig:fig1} where finite-size 
scattering region---time-dependent due to light irradiation in Fig.~\ref{fig:fig1}(a) or conventional time-independent in Fig.~\ref{fig:fig1}(b)---is attached to semi-infinite NM leads terminating at infinity into the macroscopic particle reservoirs. This is highly appropriate for Floquet TI since time-dependent potential applied in experiments~\cite{McIver2020} is confined to a finite region, either because of  a finite laser spot or the screening inside metallic contacts. On the conceptual side, such setup ensures well-defined asymptotic states and their occupation far away from the irradiated region, thereby evading technical difficulties when using time-dependent leads or reservoirs~\cite{Gaury2014}. It also ensures {\em continuous energy spectrum} of the whole system which plays a key role in both  the Landauer-B\"{u}ttiker and Kubo~\cite{Sato2019} formulation of quantum transport because it  effectively introduces dissipation at infinity and thereby steady-state transport~\cite{Perfetto2010}, while not requiring~\cite{Esin2018} to explicitly model many-body inelastic scattering processes responsible for dissipation~\cite{Imry1999}. 

However, for the same two-terminal Landauer-B\"{u}ttiker setup with irradiated scattering region a plethora of conflicting theoretical conclusions have been reached~\cite{Rudner2020}. For example, Refs.~\cite{Kitagawa2011,Gu2011} predict quantization of longitudinal dc conductance within a few percent of $2e^2/h$, while Ref.~\cite{Kundu2014} finds its anomalous suppression. To recover the quantized value, Ref.~\cite{Kundu2013} proposed an {\em ad hoc}  summation procedure over different energies in the lead. Without utilizing such ``Floquet sum rule''~\cite{Farrell2016,Yap2017,Kundu2020}, both Refs.~\cite{FoaTorres2014,Farrell2016} confirm nonquantized $G<2e^2/h$ within $\Delta_0$ gap and $G<4e^2/h$ within $\Delta_1$ gap which, however, are largely insensitive to disorder like vacancies or on-site impurities. The precise quantization could be disrupted by dc component of pumping current~\cite{Brouwer1998,Moskalets2002}, which appears~\cite{FoaTorres2014}  even at zero bias voltage  due to time-dependent potential in the Hamiltonian whenever the left-right symmetry of the device is broken statically or dynamically~\cite{FoaTorres2005,Bajpai2019}. 

The absence of quantization is explained~\cite{Esin2018,Rudner2020,FoaTorres2014,Farrell2016} by the mismatch between nonirradiated electronic states in the NM leads and edge  states within the gaps of the Floquet TI. The mismatch between states in topologically trivial NM leads and TI scattering region exists also in conventional time-independent TI devices, but without significant disruption of quantized conductance in Fig.~\ref{fig:fig2}. However, specific to  Floquet TIs is possibility of Floquet replicas to couple to bulk bands~\cite{FoaTorres2014,Fedorova2019}. That is, although edge states within the gap $\Delta_0$ are  primarily built from states near the CNP of nonirradiated graphene, they also contain harmonic components near $\pm n\hbar \omega$ which open possibility for electronic photon-assisted tunneling into or out of states in the NM leads whose energies are far away from the CNP. Thus, engineering the density of state of the leads, in order to connect Floquet TI and macroscopic reservoirs through a narrow band of filter states, can recover longitudinal dc conductance within a few percent of $2e^2/h$~\cite{Esin2018}.

The issue of experimentally detectable quantized conductance can be examined without resorting to time-independent Floquet formalism, that is, by performing direct time-dependent quantum transport simulations~\cite{Gaury2014}. Due to high computational demand, such calculations are rarely pursued, but some 
attempts yield longitudinal conductance reaching close to quantized value after sufficiently long time~\cite{Fruchart2016}. This then poses a question on the accuracy of truncation procedure that is inevitably done to reduce infinite matrices in the  Floquet formalism where artifacts~\cite{Mahfouzi2012} can be introduced. One  such artifact is dc current which is not conserved (i.e., different in the left and right lead)~\cite{Kitagawa2011,Wang2003}, or insufficient number of Floquet replicas is retained for achieving converged results.

In this study, we employ {\em charge-conserving} solution~\cite{Mahfouzi2012} for the Floquet-nonequilibrium Green functions (Floquet-NEGF)~\cite{Kitagawa2011,Wang2003,Mahfouzi2012} which ensures that dc current in the left (L) and the right (R) lead  are identical at each level of truncation of matrices in the Floquet formalism, i.e., the number of of ``photon'' excitations $N_\mathrm{ph}$ retained. As an overtire, Fig.~\ref{fig:fig3}(d) demonstrates $|I_\mathrm{L}| \equiv |I_\mathrm{R}|$ at each $N_\mathrm{ph}$, as well as that dc component of current converges at  $N_\mathrm{ph}=7$. Nevertheless, the conductance in Fig.~\ref{fig:fig3}(b) remains nonquantized in both $\Delta_0$ and $\Delta_1$ gaps. We then proceed to compare spatial profiles of local current density in conventional and Floquet TIs in Fig.~\ref{fig:fig4} which offer detailed microscopic insight on how electrons propagate from one to another carbon atom as they transition from topologically trivial NM leads into the TI region, or within the TI region with possible edge or bulk vacancies introduced as disorder. 

The paper is organized as follows. Section~\ref{sec:mm_a} describes the Hamiltonian of Floquet TI defined on ZGNR, as well as charge-conserving Floquet-NEGF from Ref.~\cite{Mahfouzi2012}, which is extended here to nonzero bias voltage and to obtain local current density. The same ZGNR is used in Sec.~\ref{sec:mm_b} to define  Hamiltonians for the conventional time-independent QHI, QAHI and QSHI, where we also provide steady-state NEGF expressions for  local current density in these systems. Section~\ref{sec:results_a} presents results for two-terminal conductance of these four TIs, and Sec.~\ref{sec:results_b} compares spatial profiles of local current density as it flows from the NM leads into those four TIs.  In Sec.~\ref{sec:exp}  we discuss experimental schemes to quantify  bulk vs. edge contributions to total current within Floquet TI using either a nanopore~\cite{Heerema2016,Chang2014,Chang2012} drilled in the interior of irradiated ZGNR, whose effect on the conductance is also explicitly calculated, or magnetic field imaging via diamond NV centers~\cite{Ku2020}. We conclude in Sec.~\ref{sec:conclusions}.

 \section{Models and Methods}\label{sec:models}
 \subsection{Hamiltonian and quantum transport formalism for Floquet TI}\label{sec:mm_a}
\begin{figure*}
	\includegraphics[width=\linewidth]{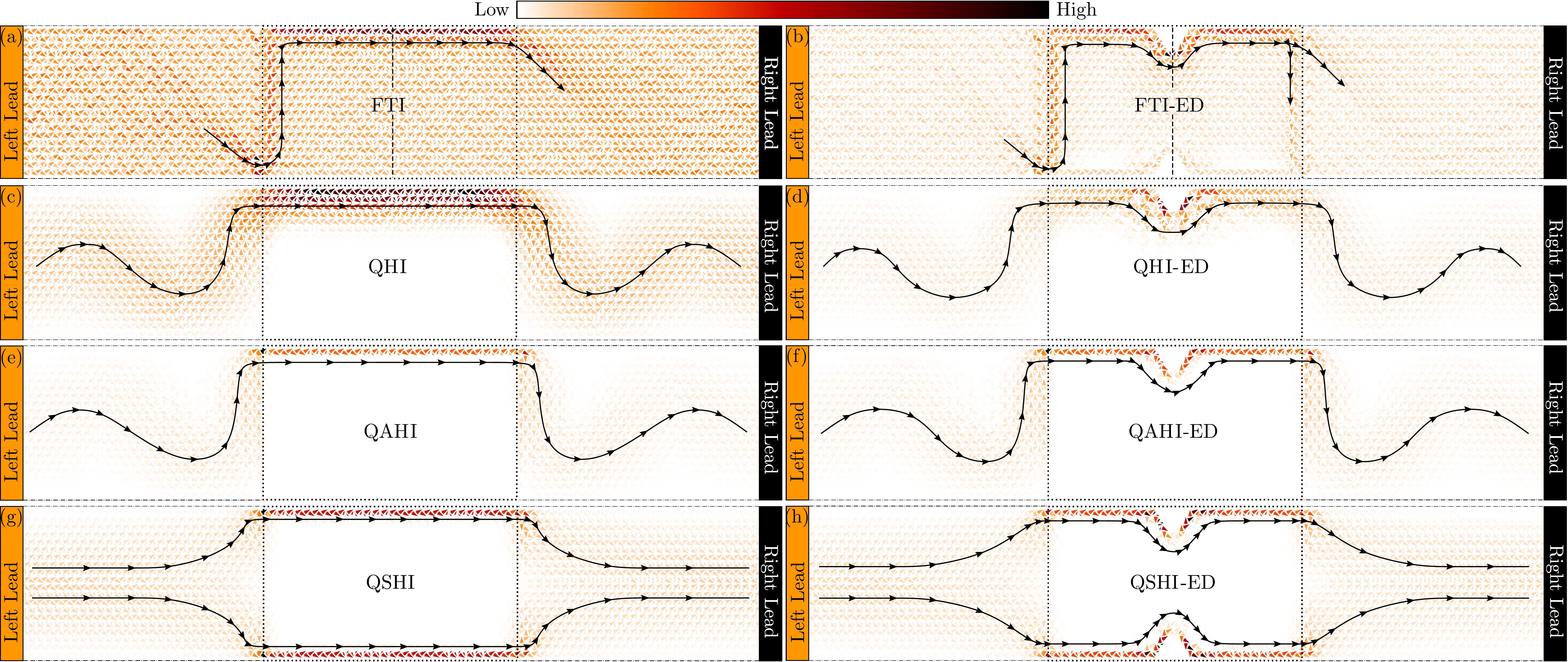}
	\caption{Spatial profiles of local current density in two-terminal devices of Fig.~\ref{fig:fig1} where the scattering region (dotted rectangle) of finite length is: (a) irradiated pristine ZGNR hosting Floquet TI; (b) irradiated edge-disordered ZGNR; (c) pristine QHI; (d) edge-disordered QHI; (e) pristine QAHI; (f) edge-disordered QAHI; (g) pristine QSHI; and (h) edge-disordered QSHI.  In panels (a) and (b)  we use $\hbar\omega = 3\gamma$, $z=0.5$, $N_\mathrm{ph} = 7$, and $E_F = \hbar \omega /2$ corresponding to the middle of $\Delta_1$ gap in Fig.~\ref{fig:fig3}(a). In panels (c)--(h), \mbox{$E_F = 0.2 \gamma$}, and in (g) and (h) \mbox{$t_\mathrm{SO} = 0.1 \gamma$}. The black solid arrows are guide to the eye to indicate the spatial region with large flux  of local current density.}
	\label{fig:fig4}
\end{figure*}

The semi-infinite leads and the scattering region in Fig.~\ref{fig:fig1} combined constitute, prior to introducing light or external magnetic field or spin-orbit coupling into the scattering region, an infinite homogeneous ZGNR described by the nearest-neighbor tight-binding Hamiltonian
\begin{equation}\label{eq:gr_ham_nir}
\hat{H}_\mathrm{ZGNR} = - \sum_{\braket{ij}} \gamma_{ij} \hat{c}^\dagger_i\hat{c}_j,
\end{equation}
Here $\braket{ij}$ indicates the sum over the nearest-neighbor sites; $\hat{c}_i^\dagger$($\hat{c}_j$) creates (annihilates) an electron on site $i$ of the honeycomb lattice hosting a single $p_z$-orbital $\langle \mathbf {r}| i \rangle=\pi(\mathbf{r}-\mathbf{R}_i)$; and \mbox{$\gamma_{ij} = \gamma = 2.7$ eV} is the  nearest-neighbor hopping from site $i$ to $j$. The width of the ZGNR is chosen as $W= 29a$, where $a$ is the distance between two  nearest-neighbor carbon atoms in graphene. It is well-known that, in general, TIs thinner than twice the width of their boundary states will experience hybridization of those states and opening of a topologically trivial mini gap~\cite{Hasan2010,Qi2011,Chang2015} at the crossing point. For Floquet TI studied in Fig.~\ref{fig:fig3}(a) this would happen if $W \le 14a$, so that our choice of $W$ evades such size artifacts. This is also ensured in the cases for QHI, QAHI and QSHI in Fig.~\ref{fig:fig4} where ZGNR is always wider than the width of edge currents. The ZGNR terminates at infinity into the macroscopic reservoirs of electrons  whose chemical potentials are $\mu_\mathrm{L} = E_{F} + eV_b/2$ and $\mu_\mathrm{R} = E_{F} - eV_b/2$ for $E_F$ as the Fermi energy and $V_b$ as the applied dc bias voltage. Note that zero-temperature two-terminal conductance $G(E_F)$ of an infinite homogeneous ZGNR described by Hamiltonian in Eq.~\eqref{eq:gr_ham_nir} is plotted for comparison in Fig.~\ref{fig:fig3}(b) and labeled as nonirradiated (NIR).

In the case of Floquet TI, circularly polarized monochromatic laser light irradiates the scattering region (shaded blue) of finite length $L = 30\sqrt{3}a$ in Fig.~\ref{fig:fig1}(a). The electromagnetic field of light is introduced into the Hamiltonian through the vector potential \mbox{$\mathbf{A}(t) =  A_0(\bold{e}_x \cos\omega t  + \bold{e}_y \sin\omega t )$}, where $\bold{e}_x$ ($\bold{e}_y$) is the unit vector along the $+x$-axis ($+y$-axis). The corresponding electric field generated by $\mathbf{A}(t)$ is \mbox{$\bold{E}(t) = -\partial \bold{A}(t)/\partial t$}. We neglect the relativistic magnetic field of light, so that electronic spin degree of freedom maintains its degeneracy and it is excluded from our analysis. The vector potential 
modifies the Hamiltonian in Eq.~\eqref{eq:gr_ham_nir} via the standard Peierls substitution~\cite{Li2020} 
\begin{equation}\label{eq:pphase}
 \hat{c}^\dagger_i \hat{c}_j \longmapsto \exp\bigg[ i2z(\bold{e}_x \cos\omega t  + \bold{e}_y \sin\omega t)\cdot \bold{r}_{ij}\bigg]\hat{c}^\dagger_i \hat{c}_j,
\end{equation}
which is rigorously proven~\cite{Panati2003} to be sufficient to capture the leading order effects due to the presence of the vector potential $\mathbf{A}(t)$. Here $z = eaA_0/2\hbar$ is a dimensionless measure of intensity of the circularly polarized light; $\omega$ is the frequency and $\bold{r}_{ij}$ is the position vector connecting site $i$ with site $j$. The new Hamiltonian $\hat{H}(t)$ with time-dependent hopping  between sites $i$ and $j$, $\gamma_{ij}(t) = \gamma \exp\bigg[ i2z(\bold{e}_x \cos\omega t  + \bold{e}_y \sin\omega t)\cdot \bold{r}_{ij}\bigg]$, is time-periodic, $\hat{H}(t+T)=\hat{H}(t)$, with period $T = 2\pi/\omega$. 

Any solution of the Schr\"{o}dinger equation, $i\hbar \partial \Psi(t)/\partial t = \hat{H}(t) \Psi(t)$, with time-periodic Hamiltonian $\hat{H}(t)=\hat{H}(t+T)$  can be expressed as a linear combination, $\Psi(t)=\sum_\alpha c_\alpha \phi_\alpha^\mathrm{F}(t)$, of the so-called Floquet functions~\cite{Shirley1965,Sambe1973}
\begin{equation}\label{eq:eigenstates}
\phi_\alpha^\mathrm{F}(t) = e^{-i \xi_\mathrm{QE}^\alpha t/\hbar} u_\alpha(t).
\end{equation}
Here  $\xi_\mathrm{QE}^\alpha$ is the Floquet {\em quasi-energy} and \mbox{$u_\alpha(t+T)=u_\alpha(t)$} are periodic functions which can, therefore, be expanded into a Fourier series
\begin{equation}\label{eq:ufunction} 
u_\alpha(\mathbf{r},t)=\sum_{n=-\infty}^{\infty} e^{i n \omega t} u_n^\alpha(\mathbf{r}).
\end{equation} 
The time-periodic Hamiltonian $\hat{H}(t)=\hat{H}(t+T)$ can also be expanded into a Fourier series
\begin{equation}\label{eq:ham_fourier}
\hat{H}(t) = \sum_{n=-\infty}^{\infty} \hat{H}_n e^{i n \omega t},
\end{equation}
where $\hat{H}_n$ is given in  terms of the Bessel functions $J_{m}(z)$ of the first kind
\begin{subequations}\label{eq:jmcs}
\begin{eqnarray}\label{eq:jms}
\exp(iz\sin x) = \sum_{m=-\infty}^{\infty} J_{m}(z)e^{i m x},
\end{eqnarray}
\begin{eqnarray}\label{eq:jmc}
\exp(iz\cos x) = \sum_{m=-\infty}^{\infty} i^m J_{m}(z)e^{i m x}.
\end{eqnarray}
\end{subequations}
Using the matrix representation of the Fourier coefficients $\bold{H}_n$ in Eq.~\eqref{eq:ham_fourier} in the basis of orbitals $|i\rangle$, we construct the  Floquet Hamiltonian~\cite{Shirley1965,Sambe1973} 
\begin{equation}\label{eq:floquet_ham}
\check{\bold{H}}_\mathrm{F} = \begin{pmatrix}
\ddots & \vdots & \vdots & \vdots & \iddots\\
\cdots & \bold{H}_0 & \bold{H}_1 & \bold{H}_2 & \cdots \\
\cdots & \bold{H}_{-1} & \bold{H}_0 & \bold{H}_1 & \cdots \\
\cdots & \bold{H}_{-2} & \bold{H}_{-1}& \bold{H}_0 & \cdots \\
\iddots & \vdots & \vdots & \vdots & \ddots
\end{pmatrix}.
\end{equation}
which is time-independent but infinite. 

The time-dependent NEGF formalism~\cite{Gaury2014} operates with two fundamental quantities~\cite{Stefanucci2013}---the retarded $\mathbf{G}^r(t,t')$ and the lesser $\mathbf{G}^<(t,t')$ Green functions (GF)---which describe the density of available quantum states and how electrons occupy those states in nonequilibrium, respectively. They depend on two times, but solutions can be sought in other representations, such as the double-time-Fourier-transformed~\cite{Mahfouzi2012,Wang2003} GFs, ${\bf G}^{r,<}(E,E')$. In the case of periodic time-dependent Hamiltonian, they must take the form~\cite{Martinez2003} 
\begin{equation}
{\bf G}^{r,<}(E,E')={\bf G}^{r,<}(E,E+n \hbar \omega)={\bf G}^{r,<}_n(E),
\end{equation}
in accord with the Floquet theorem~\cite{Shirley1965,Sambe1973}. The coupling of energies $E$ and $E+ n\hbar\omega$ ($n$ is integer) indicate  ``multiphoton'' exchange processes. In the absence of many-body (electron-electron or electron-boson)  interactions, currents can be expressed using solely the Floquet-retarded-GF $\check{\bold{G}}^r(E)$
\begin{equation}\label{eq:floquet_GF}
[E + \check{\bold{\Omega}} - \check{\bold{H}}_\mathrm{F} - \check{\bold{\Sigma}}^r(E) ]\check{\bold{G}}^r(E) = \check{\bold{1}},
\end{equation}
which is composed of ${\bf G}^r_n(E)$ submatrices along the diagonal. Here we use notation  
\begin{equation}\label{eq:omega_mtx}
\check{\bold{\Omega}} = \begin{pmatrix}
\ddots & \vdots & \vdots & \vdots & \iddots\\
\cdots & -\hbar\omega\bold{1} & 0 & 0 & \cdots \\
\cdots & 0 & 0 & 0 & \cdots \\
\cdots & 0 & 0 & \hbar\omega\bold{1}& \cdots \\
\iddots & \vdots & \vdots & \vdots & \ddots
\end{pmatrix},
\end{equation}
and $\check{\bold{\Sigma}}^r(E)$  is the retarded Floquet self-energy matrix
\begin{equation}\label{eq:floquet_self}
\check{\bold{\Sigma}}^r(E) = \begin{pmatrix}
\ddots & \vdots & \vdots & \vdots & \iddots\\
\cdots & \bold{\Sigma}^r(E-\hbar\omega) & 0 & 0 & \cdots \\
\cdots & 0 & \bold{\Sigma}^r(E)  & 0 & \cdots \\
\cdots & 0 & 0 & \bold{\Sigma}^r(E+\hbar\omega)  & \cdots \\
\iddots & \vdots & \vdots & \vdots & \ddots
\end{pmatrix},
\end{equation}
composed of the usual self-energies of the  leads~\cite{Velev2004}, $\bold{\Sigma}^r(E) = \sum_{p=\mathrm{L,R}}\bold{\Sigma}_p^r(E)$, on the diagonal. All matrices labeled as $\check{\mathbf{O}}$ are representations of operators acting in the Floquet-Sambe~\cite{Sambe1973} space, $\mathcal{H}_\mathrm{F} =  \mathcal{H}_T \otimes \mathcal{H}_e$, where $\mathcal{H}_e$ is the Hilbert space of electronic states spanned by localized orbitals $|i \rangle$ and $\mathcal{H}_T$ is the Hilbert space of periodic functions with period $T=2\pi/\omega$ spanned by orthonormal Fourier vectors $\langle t|n \rangle = \exp(i n \omega t)$. 

The charge current $I_{p}(t)$ in the lead $p=\mathrm{L,R}$ is time-dependent  due to Eq.~\eqref{eq:pphase}, and it also has periodicity   $T=2\pi/\omega$ like the Hamiltonian itself. The dc component of current, either due to pumping by time-dependent potential~\cite{Brouwer1998,Moskalets2002,FoaTorres2005,Bajpai2019} 
or due to applied bias voltage $V_b$ or both, is given by
\begin{equation}
I_{p}  = \frac{1}{T}\int_{t}^{t+T}  I_p(t') dt'.
\end{equation}
Such dc component, or time-averaged current over one period $T$, that is injected into the lead $p$ is obtained from the  following NEGF expression~\cite{Mahfouzi2012}
\begin{equation}\label{eq:floq_chj}
I_{p} = \frac{e}{2N_\mathrm{ph}}\int\limits_{-\infty}^{+\infty}dE\, \mathrm{Tr}[\check{\bold{\Gamma}}_p\check{\bold{f}}_p\check{\bold{G}}^r\check{\bold{\Gamma}}\check{\bold{G}}^a - \sum_{p=\mathrm{L,R}}\check{\bold{\Gamma}}_p\check{\bold{G}}^r\check{\bold{\Gamma}}_\alpha\check{\bold{f}}_\alpha\check{\bold{G}}^a ].
\end{equation}
In our convention, $I_p>0$ indicates that charge current is flowing into the lead. Here $\check{\bold{f}}_p$ is the Floquet Fermi matrix
\begin{equation}\label{eq:fermi_mtx}
\check{\bold{f}}_p(E) = \begin{pmatrix}
\ddots & \vdots & \vdots & \vdots & \iddots\\
\cdots & f_p(E-\hbar\omega)\bold{1} & 0 & 0 & \cdots \\
\cdots & 0 & f_p(E) & 0 & \cdots \\
\cdots & 0 & 0 & f_p(E+\hbar\omega)\bold{1}& \cdots \\
\iddots & \vdots & \vdots & \vdots & \ddots
\end{pmatrix},
\end{equation}
where $f_p(E)$ is the Fermi function of the macroscopic particle reservoir attached to  lead $p$; \mbox{$\check{\bold{\Gamma}}_p(E) = i [\check{\bold{\Sigma}}_p^r(E) - (\check{\bold{\Sigma}}_p^r(E))^\dagger]$} is the Floquet level broadening matrix; \mbox{$\check{\bold{\Gamma}}(E) = \sum_{p=\mathrm{L,R}} \check{\bold{\Gamma}}_p(E)$}; the Floquet-advanced-GF is defined as $\check{\bold{G}}^a(E) = [\check{\bold{G}}^r(E)]^\dagger$; and $\mathbf{1}$ is the unit matrix in $\mathcal{H}_e$ space. We note that Eq.~\eqref{eq:floq_chj} is generalization of the expression for charge current in Ref.~\cite{Mahfouzi2012} to include the applied bias voltage $V_b$. The linear-response two-terminal conductance  is then given by
\begin{equation} \label{eq:conductance}
G = \frac{I_\mathrm{R}}{V_b},
\end{equation}
for small applied bias voltage $eV_b \ll E_F$.

While the space $\mathcal{H}_e$ is finite-dimensional, with dimension equal to the number of sites $N_e$ within the scattering region, the space $\mathcal{H}_T$ is infinite-dimensional and has to be truncated using $|n| \le N_\mathrm{ph}$. For truncation we employ the following convergence criterion
\begin{equation}\label{eq:convergence}
\bigg| \frac{I_p(N_\mathrm{ph}) - I_p(N_\mathrm{ph}-1)}{I_p(N_\mathrm{ph}-1)}\bigg|\times 100  < \delta,
\end{equation}
where $\delta$ is the convergence tolerance. Since the operators acting in $\mathcal{H}_e$ are represented by matrices of dimension $N_e \times N_e$, the operators $\check{\mathbf{O}}$ acting on the truncated Floquet-Sambe space $\mathcal{H}_\mathrm{F}$ are represented by matrices of dimension $(2N_\mathrm{ph}+1)N_e \times (2N_\mathrm{ph}+1)N_e$. Note that the trace in Eq.~\eqref{eq:floq_chj}, $\mathrm{Tr} \equiv  \mathrm{Tr}_e \mathrm{Tr}_T$, is summing over contributions from different subspaces of $\mathcal{H}_T$ so that the denominator includes $2N_\mathrm{ph}$ to avoid double counting. The part of the trace operating in $\mathcal{H}_T$ space ensures that at each chosen truncation $N_\mathrm{ph}$ of Floquet replicas charge current is conserved, $I_\mathrm{L} = - I_\mathrm{R}$, unlike other types of  solutions~\cite{Wang2003,Kitagawa2011} of the Floquet-NEGF equations where current conservation is ensured only in the limit $N_\mathrm{ph} \rightarrow \infty$.

The bond current operator~\cite{Nikolic2006} between sites $i$ and $j$ is time-dependent due to Eq.~\eqref{eq:pphase} and it is given by~\cite{Gaury2014}
\begin{equation}\label{eq:bc_oper}
\begin{split}
\bold{J}_{ij}(t) &= \frac{e}{i\hbar} [\gamma_{ij}(t)\hat{c}_i^\dagger\hat{c}_j - \gamma_{ji}(t)\hat{c}_j^\dagger\hat{c}_i] \\
&= \sum_{n=-\infty}^{\infty} \bold{J}_{n}^{ij}e^{in\omega t}.
\end{split}
\end{equation}
We define the Floquet bond current matrix as
\begin{equation}\label{eq:floquet_curr}
\check{\bold{J}}_{ij} = \begin{pmatrix}
\ddots & \vdots & \vdots & \vdots & \iddots\\
\cdots & \bold{J}_0^{ij} & \bold{J}_{-1}^{ij} & \bold{J}_{-2}^{ij} & \cdots \\
\cdots & \bold{J}_{1}^{ij} & \bold{J}_0^{ij} & \bold{J}_{-1}^{ij} & \cdots \\
\cdots & \bold{J}_{2}^{ij} & \bold{J}_{1}^{ij}& \bold{J}_0^{ij} & \cdots \\
\iddots & \vdots & \vdots & \vdots & \ddots
\end{pmatrix},
\end{equation}
which yields nonequilibrium part~\cite{Nikolic2006} of dc bond (or local) charge current flowing between site $i$ and $j$ as
\begin{equation}\label{eq:floquet_bc}
J_{ij}^\mathrm{neq} =  \frac{1}{2\pi i}\sum_{n=-N_\mathrm{ph}}^{N_\mathrm{ph}} \int_{E_F+n\hbar\omega-eV_b/2}^{E_F+n\hbar\omega + eV_b/2} dE\, \mathrm{Tr}[\check{\bold{G}}^<(E)\check{\bold{J}}_{ij}],
\end{equation}
where $\check{\bold{G}}^<(E) = \sum_{p=\mathrm{L,R}} i\check{\bold{G}}^r(E)\check{\bold{\Gamma}}_p(E)\check{\bold{f}}_p(E)\check{\bold{G}}^a(E)$. 

\subsection{Hamiltonian and quantum transport formalism for QHI, QAHI and QSHI}\label{sec:mm_b}

The two-terminal setup in Fig.~\ref{fig:fig1}(b) hosts one of the three conventional time-independent TIs as the scattering region (shaded green) of finite length $L = 30\sqrt{3}a$. The QHI is realized by applying an external time-independent magnetic field perpendicular to ZGNR. The magnetic field is described  by a static vector potential \mbox{$\bold{A} = (By,0,0)$} in the Landau gauge, which is then included into the Hamiltonian in Eq.~\eqref{eq:gr_ham_nir} via the Peierls substitution~\cite{Li2020,Panati2003}
\begin{equation}\label{eq:qh}
 \hat{c}^\dagger_i \hat{c}_j \longmapsto \exp\bigg[ i \frac{\beta}{a_0^2}(x_i-x_j)(y_i+y_j)\bigg]\hat{c}^\dagger_i \hat{c}_j.
\end{equation}
Here $(x_i,y_i)$ indicates the position vector of a carbon atom at site $i$, and $\beta = eBa_0^2/\sqrt{3}\hbar \approx 0.07$ is a dimensionless measure of the magnetic field strength.

The QAHI~\cite{Liu2016} is described by the Haldane model~\cite{Haldane2017,Haldane1988} on the honeycomb lattice
\begin{eqnarray}\label{eq:qah}
\hat{H}_\mathrm{QAHI} & = & \sum_{\braket{ij}}- \gamma_{ij}  \hat{c}^\dagger_i\hat{c}_j + \sum_{\braket{\braket{ij}}} \widetilde{\gamma}_{ij}\hat{c}^\dagger_i\hat{c}_j  \nonumber \\
&& + \sum_{i \in A} m \hat{c}_i^\dagger\hat{c}_i + \sum_{i \in B} (-m) \hat{c}_i^\dagger\hat{c}_i.
\end{eqnarray}
Here $\braket{\braket{ij}}$ indicates the sum over the next-nearest-neighbor sites, and $\widetilde{\gamma}_{ij} = - \widetilde{\gamma}_{ji} = i \widetilde{\gamma}$ where we use $\tilde{\gamma}=0.14 \gamma$. The last two mass terms on the right hand side have different sign on the triangular sublattices A and B of the honeycomb lattice, where $m=0.2 \gamma$ specifies the ``mass'' term. Note that circularly polarized light employed in Eq.~\eqref{eq:pphase} is mandatory for Floquet TI to mimic QAHI phase of the Haldane model. In contrast, linearly polarized light, which is made of equal superposition of clockwise and anticlockwise circular polarizations, does not break time-reversal symmetry and cannot lead to Haldane ``mass'' term.

Finally, the QSHI is described by the Kane-Mele model~\cite{Kane2005}
\begin{equation}\label{eq:kane_mele_ham}
\hat{H}_\mathrm{QSHI} = \sum_{\braket{ij}}-\gamma_{ij} \bold{c}_i^\dagger \bold{c}_j + \sum_{\braket{\braket{ij}}} it_\mathrm{SO} \bold{c}_i^\dagger \bm{\sigma} \cdot (\bold{d}_{kj}\times \bold{d}_{ik} )\bold{c}_j,
\end{equation}
whose edge states crossing the topological nontrivial band gap are both chiral and spin-polarized~\cite{Hasan2010,Qi2011}. Here $\bold{c}_i^\dagger = (\hat{c}_{i\uparrow}^\dagger,\hat{c}_{j\downarrow}^\dagger)$ is a row vector of creation operators $\hat{c}^\dagger_{i\sigma}$ that create an electron on site $i$ with spin $\sigma=\uparrow, \downarrow$; $\bold{c}_i$ is the corresponding column vector of annihilation operators; $\bold{d}_{ik}$ is the unit vector pointing from site $k$ to $i$; $\bm{\sigma} = (\hat{\sigma}_x, \hat{\sigma}_y, \hat{\sigma}_z)$ is the vector of the Pauli matrices; and $t_\mathrm{SO}$ is the strength of the intrinsic spin-orbit coupling~\cite{Kane2005,Chang2014}. 

The zero-temperature two-terminal conductance $G(E_F)=G_Q \mathcal{T}(E)$ of the setup in Fig.~\ref{fig:fig1}(b) is calculated using the Landauer transmission function~\cite{Imry1999,Stefanucci2013}
\begin{equation}
\mathcal{T}(E) = \mathrm{Tr}[\bm{\Gamma}_\mathrm{R}(E)\bold{G}^r(E)\bm{\Gamma}_\mathrm{L}(E)\bold{G}^a(E)],
\end{equation}
where the conductance quantum is $G_Q=2e^2/h$ for QHI and QAHI and  $G_Q=e^2/h$ for QSHI. Here the retarded GF of the scattering region is given by \mbox{$\bold{G}^r(E) = [E - \bold{H} - \bm{\Sigma}^r(E)]^{-1}$}; the advanced GF is $\bold{G}^a(E) = [\bold{G}^r(E)]^\dagger$; and $\bold{\Gamma}_p(E) = i [\bm{\Sigma}^r_p(E) - \bm{\Sigma}^a_p(E)]$ are the level-broadening matrices. To compute the nonequilibrium bond current between sites $i$ and $j$ we use~\cite{Mahfouzi2013}
\begin{equation}\label{eq:loc_j_ss}
J_{ij}^\mathrm{neq} = \frac{eV_b}{2\pi} \Tr[\bold{G}^r(E_F)\bm{\Gamma}_\mathrm{L}(E_F)\bold{G}^a(E_F) \bold{J}_{ij}],
\end{equation}
where $\bold{J}_{ij}$ is the bond current operator in Eq.~\eqref{eq:bc_oper} but with time-independent hopping $\gamma_{ij}$.

%

\section{Results and Discussion}\label{sec:results}
\subsection{Conductance within the topological gap: FTI vs. conventional TIs }\label{sec:results_a}

By diagonalizing $\check{\mathbf{H}}_\mathrm{F}$ in Eq.~\eqref{eq:floquet_ham} for an infinite ZGNR that is periodic along the $x$-axis and irradiated by circularly polarized light over its {\em whole} length, we obtain the quasi-energy spectrum $\xi_\mathrm{QE}(k_x)$ shown in Fig.~\ref{fig:fig3}(a). The chiral edge states crossing the  light-induced gap $\Delta_0$ at $\xi_\mathrm{QE} = 0$ (shaded yellow) and $\Delta_1$ at \mbox{$\xi_\mathrm{QE} = \pm \hbar\omega/2$} (shaded red) suggest naively that upon applying small bias voltage the zero-temperature linear-response two-terminal conductance in Eq.~\eqref{eq:conductance} should be quantized: \mbox{$G(E_F)=2e^2/h$} for $E_F$ within $\Delta_0$ gap; and \mbox{$G(E_F)=4e^2/h$} for $E_F$ within $\Delta_1$ gap due to one or two spin-degenerate conduction channels provided by the edge states, respectively. This is in analogy with chiral edge states of conventional time-independent TIs and their quantized conductance in Fig.~\ref{fig:fig2}. In contrast, the average conductance in Fig.~\ref{fig:fig3}(b) is $G(E_F) \approx 0.73 \times 2e^2/h$ within $\Delta_0$ gap and $G(E_F) \approx 1.87  \times 2e^2/h$ within $\Delta_1$ gap. We emphasize that these results are not an artifact of truncation of the Floquet Hamiltonian $\check{\mathbf{H}}_\mathrm{F}$ in Eq.~\eqref{eq:floquet_ham} since current in the L  and R lead in Fig.~\ref{fig:fig3}(d) converge  at $N_\mathrm{ph}=7$ using $\delta = 1\%$ criterion in Eq.~\eqref{eq:convergence}. Also,  our Floquet-NEGF formalism~\cite{Mahfouzi2012} ensures $|I_\mathrm{L}| \equiv | I_\mathrm{R}|$ in  Fig.~\ref{fig:fig3}(d) at each chosen $N_\mathrm{ph}$. 

We additionally plot the total density  of states (DOS) $D(E) = \sum_j D_j(E)$ in Fig.~\ref{fig:fig3}(c) which is nonzero within the gaps $\Delta_0$ and $\Delta_1$ due to contributions from the local DOS (LDOS) $D_j(E)$ originating [Fig.~\ref{fig:fig6}(a)] from both edges and bulk of ZGNR. The LDOS is extracted from the Floquet-retarded-GF in Eq.~\eqref{eq:floquet_GF} using
\begin{equation}\label{eq:ldos}
D_j (E) = \frac{i}{2\pi} \bra{j} \mathrm{Tr}_T [\check{\bold{G}}^r(E) - \check{\bold{G}}^a(E)] \ket{j},
\end{equation}
where $\mathrm{Tr}_T$ is the partial trace over states in $\mathcal{H}_T$. The issue of positivity of DOS and LDOS obtained from the Floquet-retarded-GF  has been discussed extensively~\cite{Rudner2020,Uhrig2019}.

Even though Floquet TI in irradiated ZGNR does not exhibit quantized conductance plateau in Fig.~\ref{fig:fig3}(b), its conductance is largely insensitive to ED. For example, $G(E_F)$ is reduced by $\sim 2$\% within $\Delta_0$ gap and by $\sim 15$\% within $\Delta_1$ gap upon introducing edge vacancies. Nevertheless, this is still less robust than conventional time-independent TIs whose conductance within the topologically nontrivial band gap is completely insensitive to ED, as shown in Figs.~\ref{fig:fig2}(e)--(g). The disorder is introduced by removing carbon atoms on the top and bottom edges of the scattering region, as illustrated in Fig.~\ref{fig:fig4}(b), while imposing the following conditions: (\emph{i}) ED introduced in NIR ZGNR leads to complete conductance suppression  $G(E_F) \rightarrow 0$ within the same energy interval defined by $\Delta_0$ gap; ({\em ii}) ED preserves the left-right symmetry of the device, so that charge pumping is absent when the ED ZGNR is irradiated with circularly polarized  light~\cite{FoaTorres2005,Bajpai2019,FoaTorres2014} in the absence of dc bias voltage $V_b=0$. 

Note that in the case of vacancies at the edges of QSHI, our tight-binding Hamiltonian in Eq.~\eqref{eq:kane_mele_ham} does not capture possibility of formation of a localized magnetic moment at the vacancy site which requires first-principles Hamiltonians~\cite{Vannucci2020}. This opens a possibility of backscattering involving spin flip which will disrupt~\cite{Vannucci2020} (nearly) quantized conductance in Fig.~\ref{fig:fig2}(g). 

\subsection{Spatial profiles of local current density: Floquet TI vs. QHI, QAHI and QSHI}\label{sec:results_b}

The spatial profiles of local current density, i.e., bond current  $J_{ij}^\mathrm{neq}$ defined in Eqs.~\eqref{eq:floquet_bc} and ~\eqref{eq:loc_j_ss} for Floquet TI and conventional time-independent TIs, respectively, allows us to visualize how electrons transition from topologically trivial NM leads into chiral edge states within the TI region. Figures~\ref{fig:fig4}(c), ~\ref{fig:fig4}(e) and ~\ref{fig:fig4}(g) shows that bulk states contribute to current density within the leads, but current density becomes confined to narrow flux near the  edges of QHI, QAHI and QSHI. The width of the flux corresponds to spatial extent of the edge state. As expected due to chirality of edge states, current flows only along the top edge in QHI and QAHI, while in QSHI it flows on both the top and bottom edges~\cite{Nowack2013}. This is because boundaries of QSHI host a pair of spin-polarized edge states~\cite{Kane2005}, so that on the top edge  electrons with spin $\sigma=\uparrow$ move from left to right while at the bottom edge electrons with spin $\sigma=\downarrow$ move from left to right. Upon introducing ED in Figs.~\ref{fig:fig4}(d), ~\ref{fig:fig4}(f) and ~\ref{fig:fig4}(h), topological protection and quantized transport through edge states manifest by local current density circumventing the disorder since any backscattering would require to cross over to  the other edge which is forbidden due to the absence of bulk states~\cite{Buttiker1988}. 

\begin{figure}
	\includegraphics[width=\linewidth]{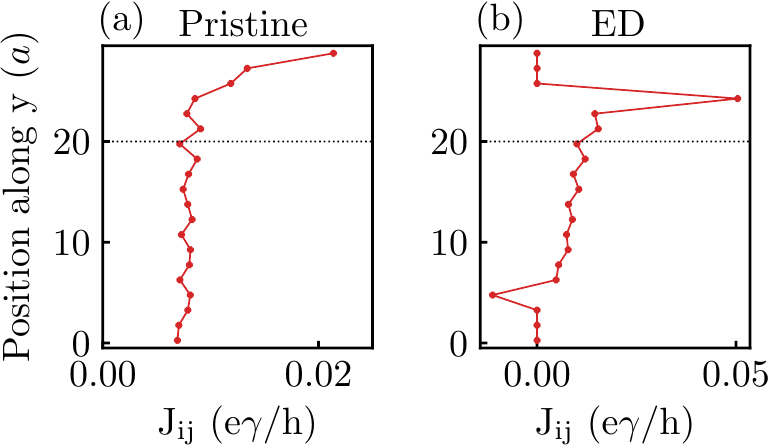}
	\caption{Spatial profile of local current density in Figs.~\ref{fig:fig4}(a) and ~\ref{fig:fig4}(b) over the transverse cross section within: (a) pristine Floquet TI; or (b) Floquet TI with edge disorder. The position of the transverse cross section is marked by dashed vertical line in Figs.~\ref{fig:fig4}(a) and ~\ref{fig:fig4}(b), respectively. The horizontal dashed line in both panels marks the extent of the edge state.}
	\label{fig:fig5}
\end{figure}

In contrast, local current density is nonzero within the whole Floquet TI in Fig.~\ref{fig:fig4}(a), with larger flux near the edges [Fig.~\ref{fig:fig5}(a)]. Upon introducing ED, the edge flux circumvents the disorder but due to general nonlocality of quantum transport bulk flux is also reduced [Fig.~\ref{fig:fig5}(b)] which explains slight reduction of conductance in Fig.~\ref{fig:fig3}(b) within gaps $\Delta_0$ and $\Delta_1$. 

Interestingly, SQUID-based imaging of QSHI made from HgTe quantum wells has found that gate tuning of bulk conductivity can lead to transport regime where edge and bulk local current densities {\em coexist}~\cite{Nowack2013}. The trace of local current density is scanned by detecting its magnetic field produced according to the  Biot-Savart law, which is possible even through the top gate employed to tune the carrier density. In this regime, experimental images were analyzed to quantify contribution of edge and bulk local currents to the total current. We perform similar analysis in Fig.~\ref{fig:fig5} which shows that in pristine Floquet TI from Fig.~\ref{fig:fig4}(a), edge current contributes 44\% and bulk current contributes 56\% to the total current over the transverse cross section [marked by dashed line in Fig.~\ref{fig:fig4}(a)]. Conversely, in the presence of edge disorder in  Fig.~\ref{fig:fig4}(b), edge current contributes 52\% and bulk current contributes 48\% to the total current over  the same transverse cross section.

\subsection{Proposed experimental schemes for probing edge vs. bulk transport within Floquet TI: Graphene nanopore and magnetic field imaging}\label{sec:exp}

\begin{figure}
	\includegraphics[width=8.5cm]{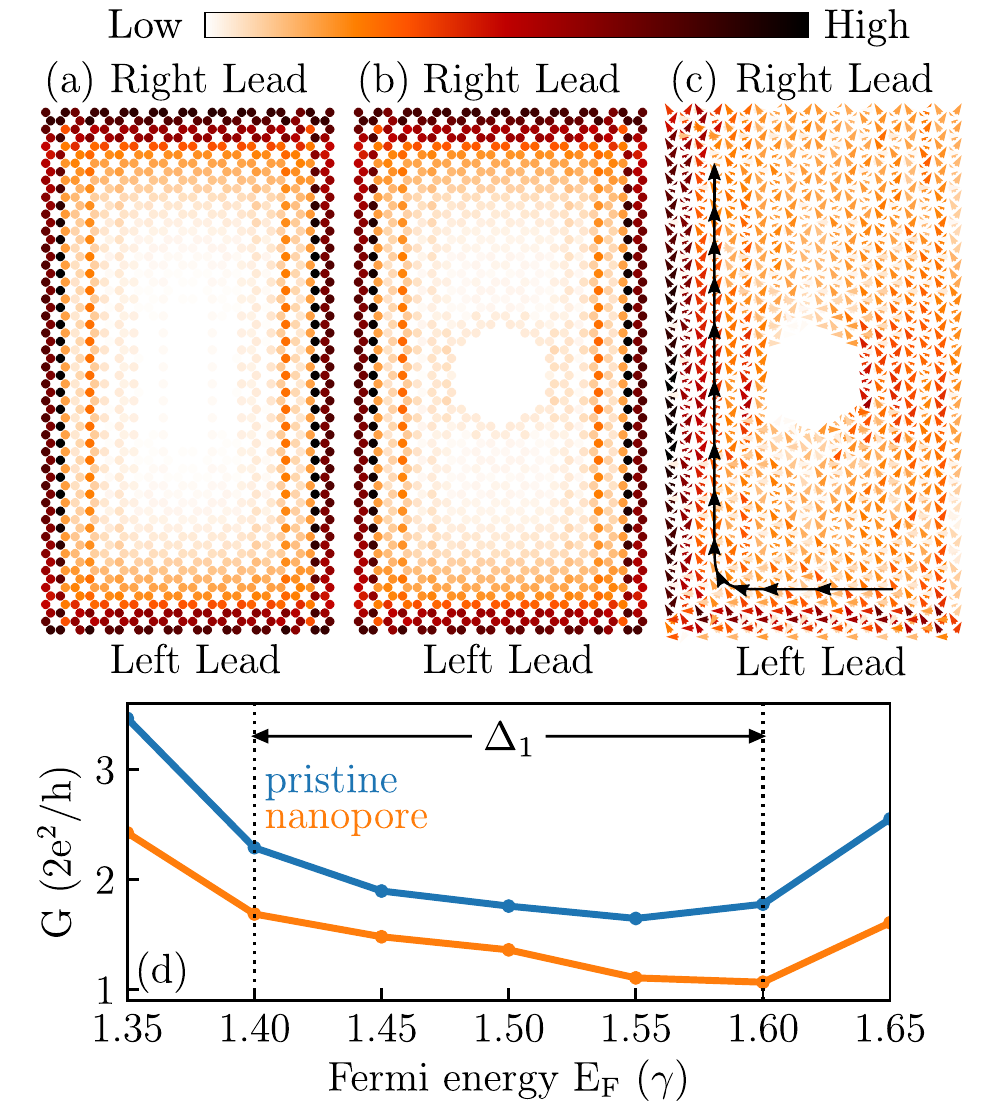}
	\caption{(a) The LDOS [Eq.~\eqref{eq:ldos}] evaluated at $E = \hbar\omega/2$ in the center of $\Delta_1$ gap in Fig.~\ref{fig:fig3}(a) for irradiated pristine ZGNR. (b) The LDOS evaluated at $E=\hbar\omega/2$  for irradiated ZGNR with a nanopore drilled in the interior of the nanoribbon. (c) Time-averaged local bond current [Eq.~\eqref{eq:floquet_bc}] in the same irradiated ZGNR with a nanopore as in (b). (d) Zero-temperature two-terminal conductance $G(E_F)$ of irradiated ZGNR with a nanopore (orange line) vs. conductance of irradiated pristine ZGNR (blue line) within the gap $\Delta_1$ in Fig.~\ref{fig:fig3}(a). The former is reduced by $\sim 28\%$ with respect to the latter. In all panels we use $\hbar\omega = 3\gamma$, $z=0.5$ and $N_\mathrm{ph} = 7$.}
	\label{fig:fig6}
\end{figure}

The spatial profiles of local current density of conventional time-independent TIs in Figs.~\ref{fig:fig4}(c)--(h) indicate that any disorder introduced in the interior of ZGNR will have no effect on the two-terminal conductance in Fig.~\ref{fig:fig2}. This was explicitly demonstrated in Ref.~\cite{Chang2014} for the case of QSHI (based on graphene plus heavy adatoms). Therefore, we propose to employ a nanopore in the ZGNR interior as the simplest technique that can detect the presence of bulk current density in Figs.~\ref{fig:fig4}(a) and ~\ref{fig:fig4}(b) in the case of Floquet TI. We introduce nanopore in Figs.~\ref{fig:fig6}(b) and ~\ref{fig:fig6}(c) in  such a way that it preserves the left-right symmetry of the device in order to avoid any charge pumping by time-dependent potential of light~\cite{FoaTorres2005,Bajpai2019,FoaTorres2014}. In experiments, nanopores are routinely drilled, without disrupting the surrounding honeycomb lattice of graphene, for applications like DNA sequencing~\cite{Heerema2016}, and they could also be deployed to block phonon transport in thermoelectric applications~\cite{Chang2014,Chang2012}.  Figures~\ref{fig:fig6}(a) and ~\ref{fig:fig6}(b) confirms that nanopore does not impair high LDOS [Eq.~\eqref{eq:ldos}] near the edges of the Floquet TI, which correspond to chiral edge states from Fig.~\ref{fig:fig3}(a).  Figure~\ref{fig:fig6}(c) shows that local transport in the presence of nanopore utilizes {\em both} left-to-right moving chiral edge states and bulk states. Since electrons flowing through the bulk  are backscattered by the nanopore, presence of nanopore reduces conductance by about $\simeq 28$\% in Fig.~\ref{fig:fig6}(d) within  the gap $\Delta_1$.

A more detailed probing of edge vs. bulk transport in  $\sim\mu$m-sized devices, such as those employed in recent experiments~\cite{McIver2020} to convert graphene into Floquet TI, could be achieved using diamond NV centers. The device can be fabricated on a diamond containing high-density, near-surface NV  ensemble~\cite{Tetienne2017,Ku2020}, along with a graphite top gate separated by hexagonal boron nitride to tune the carrier density~\cite{Ku2020}. The spin state of NV centers, which serve as the sensor of magnetic field produced by local current density, can be optically initialized and readout via imaging the NV photoluminescence onto a camera. Such a setup has the advantages of being able to operate over a wide range of temperatures, from cryogenic to room temperature (e.g., the experiment in Ref.~\cite{McIver2020} was done at 80 K); it can be readily integrated with an optical cryostat necessary for experiments involving THz radiation; and it has less stringent vibrational requirement compared to scanning setups. We note that THz radiation is far detuned from any of the NV orbital/spin transitions and hence it will not affect NV centers at all.

In Ref.~\cite{McIver2020}, a constant DC bias generates a current \mbox{$I \simeq 125$ $\mu$A}, and THz pulses drive the system into Floquet TI state for about \mbox{$3$ ps} at \mbox{$\sim 210$ kHz} repetition rate. Hence, one has a time-averaged typical current density \mbox{$\bar{J}_{\mathrm{F}}\sim 80$ pA/$\mu$m} in a 1-$\mu$m-wide device. This corresponds to a typical stray magnetic field \mbox{$\mu_0 \bar{J}\sim 0.1$ nT}, where $\mu_0$ is the permeability of free space. While it is a small field, its measurement is attainable with existing NV sensing technologies. For example, a single NV can sense $\sim$ nT field with a Hahn-echo sequence over 100 seconds signal averaging at room temperature~\cite{Maze2008}. Detection of \mbox{$\sim 0.1$ nT} field is attainable in combination with entanglement-assisted repetitive readout~\cite{Neumann2010,Lovchinsky2016}, which is available for NV ensemble, as well as with enhanced coherence at cryogenic temperatures and with dynamical decoupling sequences~\cite{BarGill2013}. One can measure the differential current density \mbox{$\Delta J(x,y)\equiv J_{\mathrm{FTI}}(x,y)-J_{\mathrm{NIR}}(x,y)$}, where $J_{\mathrm{FTI}}(x,y)$ [$J_{\mathrm{NIR}}(x,y)$] is current density within the Floquet TI [nonirradiated normal phase], by pulsing on the THz radiation during one free precession time of the Hahn-echo and keeping the THz drive off during the other free precession. The current density $J_{\mathrm{NIR}}(x,y)$ can be measured separately in a Hahn-echo measurement without any THz pulses to enable one to extract $J_{\mathrm{FTI}}(x,y)$. Diffraction-limited optical imaging of magnetic field has resolution $\sim$ 400 nm~\cite{Ku2020}, which is enough to resolve edge currents separated by a width of 1 $\mu$m. With further improvement in spatial resolution, we anticipate that $\sim$ 10 nm resolution can be achieved by using the Fourier gradient imaging~\cite{Arai2015}.

\section{Conclusions}\label{sec:conclusions}

In conclusion, using steady-state NEGF formalism applied to two-terminal Landauer-B\"{u}ttiker setup [Fig.~\ref{fig:fig1}(b)] with scattering region consisting of conventional time-independent TIs---such as QHI, QAHI and QSHI defined on graphene nanoribbon in order to generate chiral edge states of {\em finite} length---we demonstrate that their conductance is never perfectly quantized [Fig.~\ref{fig:fig2}]. This is  due to backscattering at the NM-lead/2D-TI interface. Nevertheless, it remains very close to perfect plateau at $2e^2/h$ within the topologically nontrivial band gap, and it is completely insensitive to edge disorder. The spatial profiles of local current density visualize how electrons flow from bulk states within topologically trivial NM leads into the narrow flux defined by edge states within the TI region, while circumventing any edge disorder within the TI region.  

In contrast, when the scattering region is converted into the Floquet TI by irradiating graphene nanoribbon by circularly polarized light, conductance within light-induced topologically nontrivial band gaps is not quantized, but it changes little with edge disorder. This results confirm previous findings in the literature~\cite{FoaTorres2014,Farrell2016} while ensuring proper convergence and charge current conservation in the solution of Floquet-NEGF equations~\cite{Mahfouzi2012}. Furthermore, we employ such charge-conserving  Floquet-NEGF formalism to compute spatial profiles of local current density. They are higher along the edges [Fig.~\ref{fig:fig5}(a)], following high LDOS near the edges  [Fig.~\ref{fig:fig6}(a)], but they remains nonzero also in the interior of the Floquet TI [Fig.~\ref{fig:fig4}(a)]. Such spatial profiles make it also possible to refine  previous qualitative estimates of edge vs. bulk contribution to current through Floquet TI~\cite{Esin2018} with precise measure from Figs.~\ref{fig:fig4}(a), ~\ref{fig:fig4}(b) and ~\ref{fig:fig5} which show that edge currents and bulk currents contribute nearly equally to the total current. Thus, observing quantized transport in Floquet TI would require to minimize coupling to bulk states. 

We propose a very simple experimental technique to detect presence or absence of bulk states in quantum transport through Floquet TI---conductance measurements under laser irradiation should be performed using uniform graphene flake, as in the very recent experiments~\cite{McIver2020}, as well as repeated after a nanopore~\cite{Heerema2016} is drilled in the interior of the flake. If local current density is nonzero in the bulk, it will be scattered by the nanopore which leads to $\simeq 28$\% reduction [Fig.~\ref{fig:fig6}(d)] of the two-terminal conductance when compared to graphene nanoribbon without the nanopore. Finally, we delineate more sophisticated experimental schemes for direct imaging~\cite{Ku2020} of magnetic field produced by edge and bulk local current densities based on diamond NV centers whose orbital/spin transitions are far detuned from THz radiation employed~\cite{McIver2020} in recent experiments to convert graphene into Floquet TI.

\begin{acknowledgments}
We thank L. E. F. Foa Torres, J. Gong and L. Vannucci  for insightful discussions. This research was supported by the US National Science Foundation (NSF) under Grant No. ECCS 1922689. 
\end{acknowledgments}


\begin{thebibliography}{10}
\bibitem{footnote}
For easy navigation, we provide a list of abbreviations used in the article: 2D--two-dimensional; CNP--charge neutral point; DOS--density of states; ED--edge disorder; LDOS--local density of states;  NEGF--nonequilibrium Green function; NIR--nonirradiated; NM--normal metal; NV--nitrogen-vacancy; QAHI--quantum anomalous Hall insulator; QHI--quantum Hall insulator; QSHI--quantum spin Hall insulator; SQUID--scanning superconducting quantum interference device; TI--topological insulator; ZGNR--zigzag graphene nanoribbon.
\end{thebibliography}

\end{document}